\documentclass[12pt]{article}
\usepackage{hyperref}
\usepackage{cite}
\usepackage{color}
\usepackage{graphicx}
\usepackage{amsmath}
\usepackage{amssymb}
\usepackage{xspace}

\makeatletter
\@addtoreset{equation}{section}

\makeatletter
\renewcommand\section{\@startsection {section}{1}{\z@}%
                                   {-3.5ex \@plus -1ex \@minus -.2ex}
                                   {2.3ex \@plus.2ex}%
                                   {\normalfont\large\bfseries}}
\renewcommand\subsection{\@startsection{subsection}{2}{\z@}%
                                     {-3.25ex\@plus -1ex \@minus -.2ex}%
                                     {1.5ex \@plus .2ex}%
                                     {\normalfont\bfseries}}

\def\baselinestretch{1.2}
\newcommand{\be}{\begin{equation}}
\newcommand{\ee}{\end{equation}}
\newcommand{\beq}{\begin{eqnarray}}
\newcommand{\eeq}{\end{eqnarray}}

\newcommand{\gone}[1]{{}}

\begin{document}
\begin{titlepage}
\begin{flushright}
MAD-TH-15-04
\end{flushright}

\vfil

\begin{center}

{\bf \Large
Fencing in the Swampland:\\ Quantum Gravity Constraints on Large Field Inflation}

\vfil

Jon Brown, William Cottrell, Gary Shiu and Pablo Soler

\vfil

Department of Physics, University of Wisconsin, Madison, WI 53706, USA

\vfil

\end{center}

\begin{abstract}
\noindent 
In this note we show that models of natural inflation based on closed string axions are 
incompatible with the weak gravity conjecture (WGC).  Specifically, we use T-duality in order to map the bounds on the charge-to-mass ratio of particles imposed by the WGC, to constraints on the ratio between instanton actions and axion decay constants. We use this connection to prove that if the WGC holds, even when multiple axions are present and mix with each other, one cannot have large axion decay constants while remaining in a regime of perturbative control.  We also discuss the extension of the WGC to discrete symmetries and its possible impact on models with axion monodromy, and the distinction between the strong and mild versions of the WGC. 
Finally, we offer some speculations regarding the import of these results to the general theory of inflation.

\end{abstract}
\vspace{0.5in}

\end{titlepage}
\renewcommand{\baselinestretch}{1.05}  

\section{Introduction}

Inflation is the leading theory solving a host of cosmological problems in the early universe, and it provides a seed for the observed large scale structure.  Several inflationary models, such as chaotic \cite{Linde1986395} and natural \cite{Freese:1990rb} inflation, require the inflaton to traverse a large distance (in Planck units) in field space in order to produce the requisite number of e-folds.  In the past decade, many such large field inflation models have been explored within string theory \footnote{The ultraviolet sensitivity of  large field inflation motivates one to consider inflation within string theory. For a pedagogical review of string inflation, see  e.g., \cite{Baumann:2014nda}.}. Guaranteeing a large field range presents problems of its own as Planck suppressed operators generically become sizable and thus the effective field theory  would a priori lose perturbative control.  The most popular solution to this problem is to consider axions as inflaton candidates since they enjoy a perturbative shift symmetry which suppresses quantum corrections.  Instanton effects may then generate a sufficiently small potential while breaking the continuous shift symmetry to a discrete subgroup.  Modding the axion moduli space by this subgroup, we find that the axion has a finite field range whose `radius' is known as the `axion decay constant'.  This decay constant will then set a bound on the effective field range of inflation.

From the perspective of effective field theory (EFT), requiring a large decay constant is simply a matter of tuning.  However, for field values near the Planck scale, the theory's embedding in a complete theory of quantum gravity becomes a relevant question.  Banks et al. \cite{Banks:2003sx} argued via examples that, in stringy models, single-field large axion decay constants arise \textit{only} when control of higher order instantons is lost and new light states appear in the theory.  Thus, the form of the EFT axion potential is obscured.  Most efforts in constructing large field inflation models have therefore been focused on evading this constraint by a variety of EFT mechanisms involving multiple axions \cite{Choi:2014rja, Kim:2004rp, Dimopoulos:2005ac}.  The idea is essentially that while each axion decay constant must be sub-Planckian, the net effective decay constant may be larger, as in the alignment \cite{Kim:2004rp} or N-flation \cite{Dimopoulos:2005ac} scenarios.  

Unfortunately, it is often very difficult to test directly whether these models have stringy embeddings, partially due to moduli stabilization issues.  There have been several attempts \cite{Long:2014dta, Shiu:2015uva,Shiu:2015xda, Bachlechner:2014gfa}  to engineer string realizations of the alignment scenario.  In \cite{Long:2014dta, Shiu:2015xda}, for instance the role of the axion is played by the flux through a D-brane wrapping a cycle in the internal manifold.  In general, one would expect that the tension of the D-brane shrinks the cycle it wraps and thus reduces the action of instantons wrapping the brane. This could very well lead to a loss of control as intimated by \cite{Banks:2003sx}.  One needs a mechanism to stabilize the cycle's contraction but incorporating such effects systematically in an inflationary background seems intractable. 
It was argued in 
Bachlechner et al. \cite{Bachlechner:2014gfa} 
that a `large' axion field range can be obtained
via multiple closed string axions using the semi-explicit construction of \cite{Denef:2005mm}.  In fact, \cite{Bachlechner:2014gfa} calculated a field range of $1.13 M_{p}$ while remaining (marginally) within the regime of perturbative control.  Obviously, such field ranges are still  insufficient for inflation, but the proposed model, if correct, would provide a proof of existence of models with trans-Planckian decay constants in string theory in a regime of perturbative control. We will argue in this work that such a possibility is in conflict with general properties of quantum gravity. According to our results, we predict that a more careful analysis of the stabilization process, e.g. $\alpha'$ and instanton corrections, or one-loop determinants will \textit{either} lower the effective field range \textit{or} increase the size of the instanton corrections. In anticipation of our results, we will obtain a precise numerical upper bound on the inflaton field range as
our analysis does not allow for an $\mathcal{O}(1)$ fudge factor.    

The issue of determining when an EFT has a string embedding is precisely the problem of defining the `swampland' \cite{Vafa:2005ui, Ooguri:2006in}.  It would be very useful to have tools which allow one to tell, at a glance, whether a given EFT makes sense within the context of quantum gravity.  One such criteria was proposed in \cite{ArkaniHamed:2006dz}, where it was conjectured that in any UV embeddable EFT `gravity is the weakest force'.  This is known as the `Weak Gravity Conjecture' (WGC).  For example, electric repulsion always beats gravitational attraction between two electrons.  In \cite{Cheung:2014vva} an extension of this conjecture to theories with many $U(1)$'s and a general spectrum of charged particles was provided.  The extension simply states that the convex hull of a certain charge matrix must contain the `unit ball' \footnote{The precise formulation will be given later.}.  

The main purpose of this work is to relate the WGC to a condition on the axion decay constant via T-duality, and the convex hull condition of \cite{Cheung:2014vva} to models with multiple axions, and to precisify the assumptions under which natural inflation is excluded.  Specifically, we will show that if the WGC holds, even when mechanisms like kinetic mixing, alignment, and large N enhancement are considered, one cannot get a large field range while staying in a regime of perturbative control.  Moreover, we will argue that the WGC implies precise numerical bounds on inflation models rather than fuzzy statistical bounds.  These results apply to any model of axion inflation based on closed string axions.  

An interesting exception could arise if we were to somehow suppress higher instanton corrections.  One mechanism for doing this is to consider axions defined on torsional cycles.  In fact, this line of thinking leads one directly to the F-term monodromy model considered in \cite{Marchesano:2014mla}.  However, we will also discuss a possible generalization of the WGC to massive p-form fields, and argue that this model can be constrained as well.  

This paper is structured as follows.  In Section 2 we review the basics of the weak gravity conjecture and describe its formalization in terms of the convex hull condition.  Section 3 describes the T-duality between the usual WGC and the WGC as applied to axions, thus demonstrating that the conjecture is `covariant' in an appropriate sense and also fixing the precise size of the convex hull.  Section 4 proves that the convex hull condition does indeed prohibit natural inflation models in general (a caveat to our results will be discussed).   Section 5 discusses an interesting class of models that falls outside our proof but nevertheless raise issues relevant to understanding the WGC.  Next, in Section 6 we point out a general tension between inflation models and holography.  Finally, we offer our conclusions.   

The results of~\cite{Montero,Rudelius:2015xta}, which appeared while this paper was being finalized, contain a significant overlap with our results which, were presented at the MCTP Workshop ``String/M-theory Compactifications and Moduli Stabilization".

\section{The Weak Gravity Conjecture}\label{sec:WGC}

In this section we review the original formulation of the WGC and some of the arguments that support it~\cite{ArkaniHamed:2006dz}. We also describe, along the lines of~\cite{Cheung:2014vva}, its generalization to the case in which there are multiple gauge fields, and furthermore speculate about the possibility of applying the WGC to discrete gauge symmetries. 

\subsection{The conjecture}\label{sub:conjectures}
The WGC simply states that gravity should be the weakest force in any consistent theory that includes quantum gravity. Stated another way, this just says that there should be a lower bound on the strength of gauge interactions.  The simplest argument to understand this is to consider a $U(1)$ symmetry with coupling constant $g$, and take $g\to0$. In that limit, the gauge bosons decouple and the symmetry becomes global, running afoul of  `folk theorems' of quantum gravity. As is well known, in theories with global continuous symmetries one can construct black holes with an arbitrary charge that cannot be radiated away, hence leading to an infinite number of remnants, in violation of the Covariant Entropy Bound (CEB)~\cite{Susskind:1995da}.

A more precise formulation of the WGC is that for each long range force there should exist at least one particle for which the gravitational attraction is compensated by a stronger gauge repulsion. Following~\cite{ArkaniHamed:2006dz}, let us consider a $U(1)$ gauge theory with a single family of particles with mass $\mu$ and charge $e$ which does not satisfy the WGC, i.e. for which the electric repulsion is overwhelmed by the gravitational attraction. In that case, one can form perfectly stable bound states of an arbitrary number $N$ of particles with charge $eN$ and a mass $M<N\mu$ (the difference being the binding energy of the system). For large $N$ these objects will form a stable black hole. Asymptotically, in the limit $N\to \infty$, the black hole will be extremal. This infinite number of safely bound perfectly stable states leads again to a violation of the CEB.

The way to overcome this problem is to postulate that there must be a particle  that allows the infinite series of stable states to decay. The charge to mass ratio of such a particle must then be greater or equal to that of an extremal black hole
\begin{equation}\label{eq:WGC}
\frac{q}{m}\geq \frac{Q_{\small{\text{EBH}}}}{M_{\small{\text{EBH}}}}
\end{equation}

Now, there are two versions of the WGC, depending on which state is required to satisfy the condition~\eqref{eq:WGC}. 
The {\it strong} version (strong-WGC) requires that it must be the lightest charged particle in the spectrum. The {\it mild} version (mild-WGC) puts no extra condition on the state that satisfies the inequality. 
In fact, mild-WGC can be rephrased as the statement that there should be only a finite number of exactly stable and safely bound states in the theory. In the presence of multiple $U(1)$  symmetries, the requirement applies along each direction in charge space.
 
It is clear that the evidence for the strong-WGC in terms of the CEB are not as solid as those for the mild version. Nevertheless, as already mentioned in~\cite{ArkaniHamed:2006dz}, there is no known counterexample to strong-WGC in string theory.\footnote{With the possible exception of \cite{Long:2014dta}, to be discussed later, which is in disagreement with both the mild ant the strong-WGC.} In section {\ref{tension} we will argue that the covariant entropy bound does indeed provide support for the strong-WGC, however, it would be nice if one could produce some more direct and independent evidence.

\subsection{The Convex Hull Condition}
We follow in this section the approach of~\cite{Cheung:2014vva} in order to understand the consequences of the WGC in setups in which there are several $U(1)$ symmetries. For each particle $i$ with charges $\vec{q}_i$ and mass $m_i$ in the system, it is convenient to introduce the vector $\vec{z}_i=\vec{q}_i\frac{M_p}{m_i}$. In the conventions described in Section \ref{blackhole}, these vectors are normalized such that for extremal black holes\footnote{The numerical constant, $|\vec{Z}_{\text{EBH}}|$, will change depending on the space-time dimension, as discussed in \ref{blackhole}.} in 4d $|\vec{Z}|_{\text{EBH}}\equiv 1$.  Hence, for extremal black holes, $\vec{Z}$ lies on the unit sphere, while for non extremal ones $\vec{Z}$ lies inside the unit ball.  (See figure~\ref{fig:convexhull}.)

The WGC condition, which states that extremal black holes should be at most marginally stable, translates into a condition on the vectors $\vec{z}_i$ of the elementary particles in the spectrum. 
Consider an extremal black hole with charge $\vec{Q}$, mass $M$ and vector $\vec{Z}=\vec{Q}\frac{M_p}{M}$, which by definition lies on the unit sphere $|\vec{Z}|=1$. The (in)stability condition implies that there must exist a combination of $n_i$ particles of species $i$ into which the black hole can decay, i.e. $\vec{Q}=\sum n_i\vec{q}_i$ and $M\geq n_i m_i$. Define the mass fractions $\sigma_i=n_im_i/M$.  Energy and charge conservation imply $\sum_i \sigma_i\leq 1$ and $\vec{Z}=\sum_i\sigma_i \vec{z}_i$.  In other words, $\vec{Z}$ is a subunitary weighted average of $\vec{z}_i$. Since extremal black holes can be created along any direction on charge space, the condition implies that the unit ball should be contained in the convex hull spanned by the vectors $\pm \vec{z}_i$ of the system (see figure~\ref{fig:convexhull}).

The mild-WGC states that along any given direction in charge space, there is {\it some} combination of particles with total mass $m$ and total charge $\vec{q}$ whose vector $\vec{z}=\vec{q}\,\frac{M_p}{m}$ lies outside of the unit ball. It is natural to generalize the strong-WGC by requiring that the lightest (possibly multi-particle) state in any given direction in charge space satisfies the same condition, that is, $|\vec{z}_{lightest}|\geq 1$. 
Unless otherwise stated (see section~\ref{loophole}) our discussion applies to the mild-WGC, and hence automatically to the strong-WGC as well.

In figure~\ref{fig:convexhullviolation} we show several examples that fail to satisfy the convex hull condition in some direction of charge space, even if there are several states for which $|\vec{z}|\geq 1$.

\begin{figure}[tbp]
\centering
\includegraphics[width=200pt]{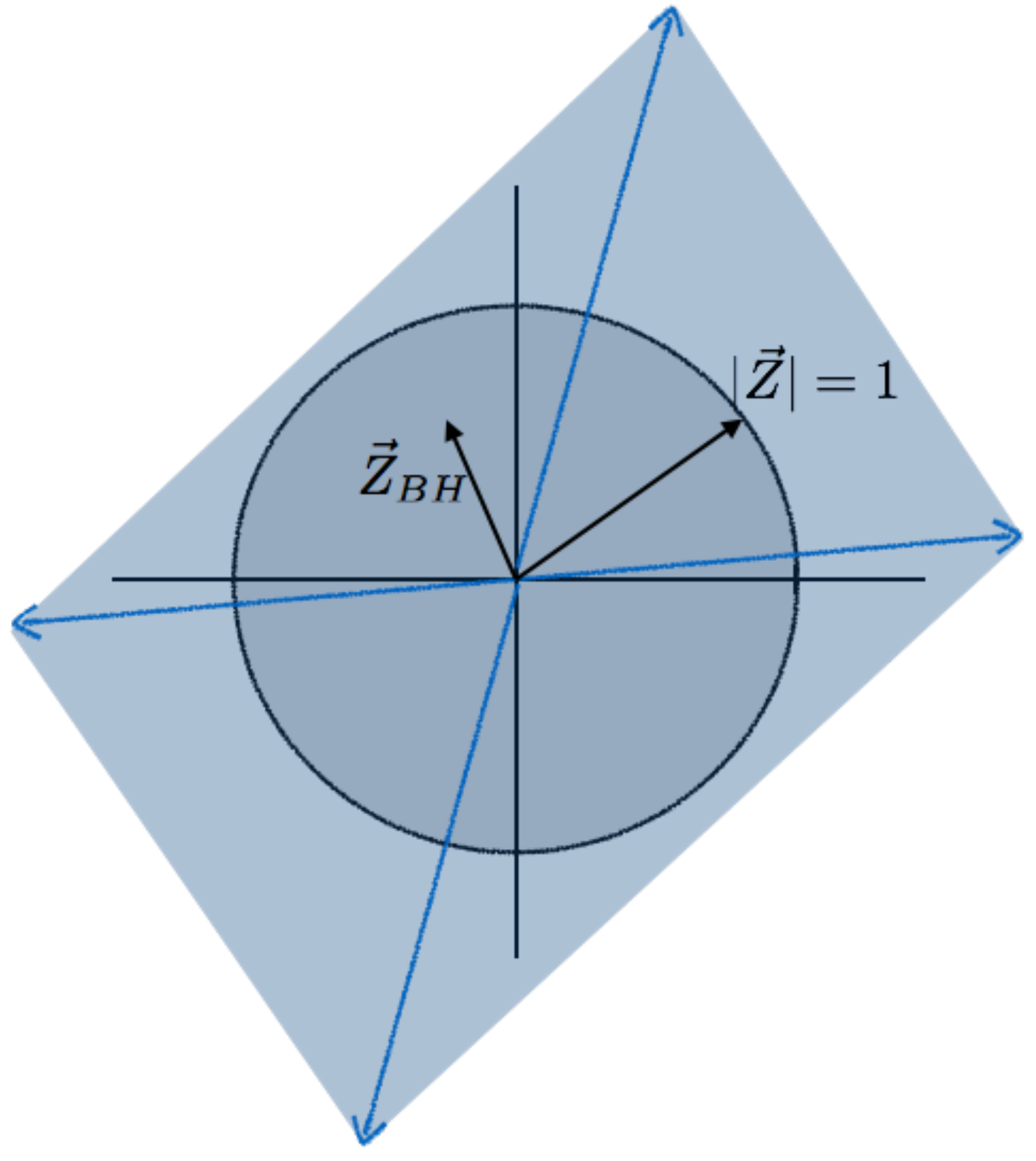}
\hspace{2cm}
\includegraphics[width=200pt]{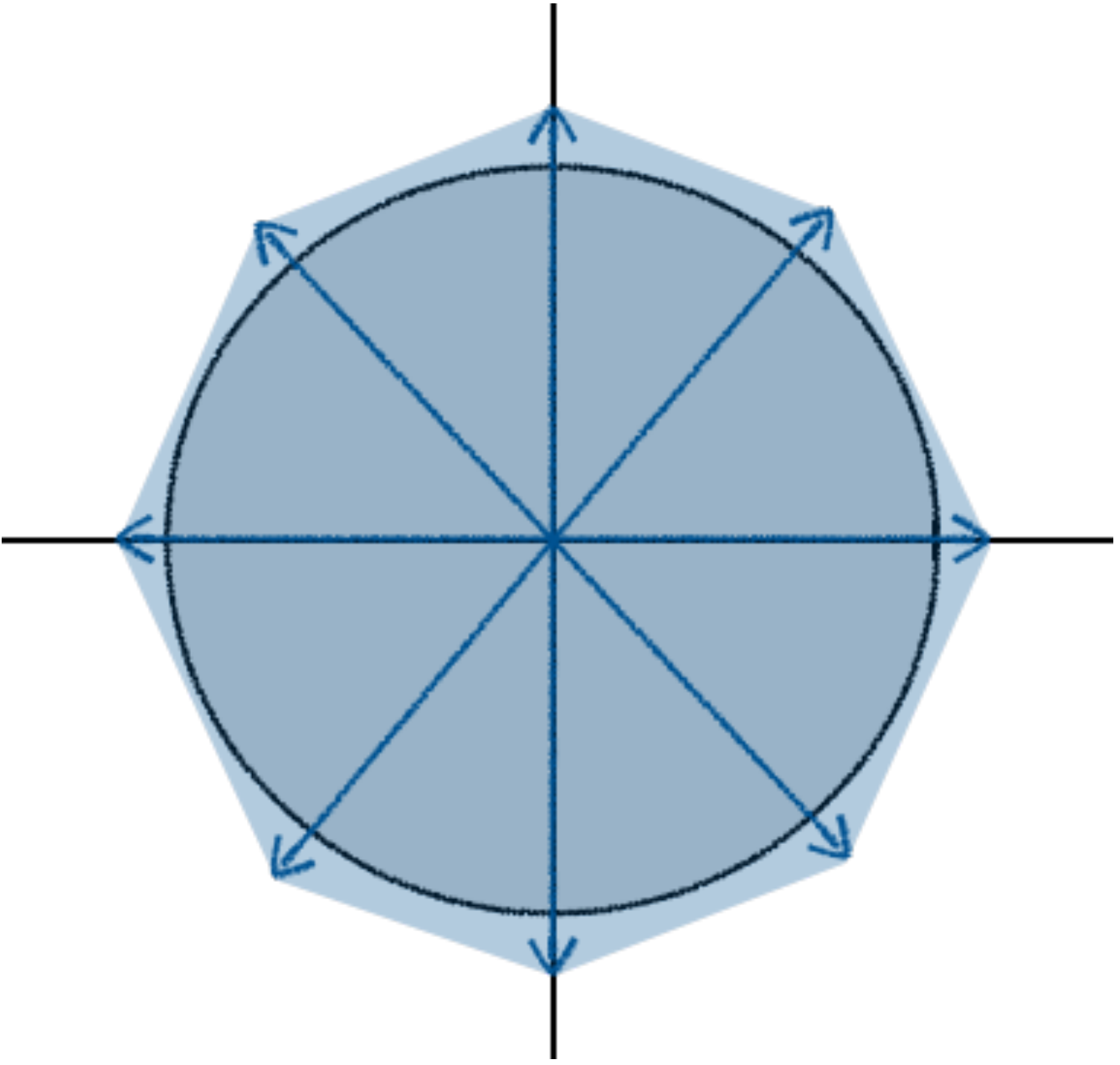}
\caption{\small The convex hull condition as applied to two $U(1)$ groups. Vectors corresponding to extremal black holes lie on the unit circle, while non-extremal black holes lie inside the unit disc. On the left (right) hand side, the convex hull defined by the $\vec{z}$ vectors of two (three) charged particles and their antiparticles (blue arrows) ensure the WGC is satisfied.}\label{fig:convexhull}
\end{figure}

\begin{figure}[htbp]
\centering
\includegraphics[width=175pt]{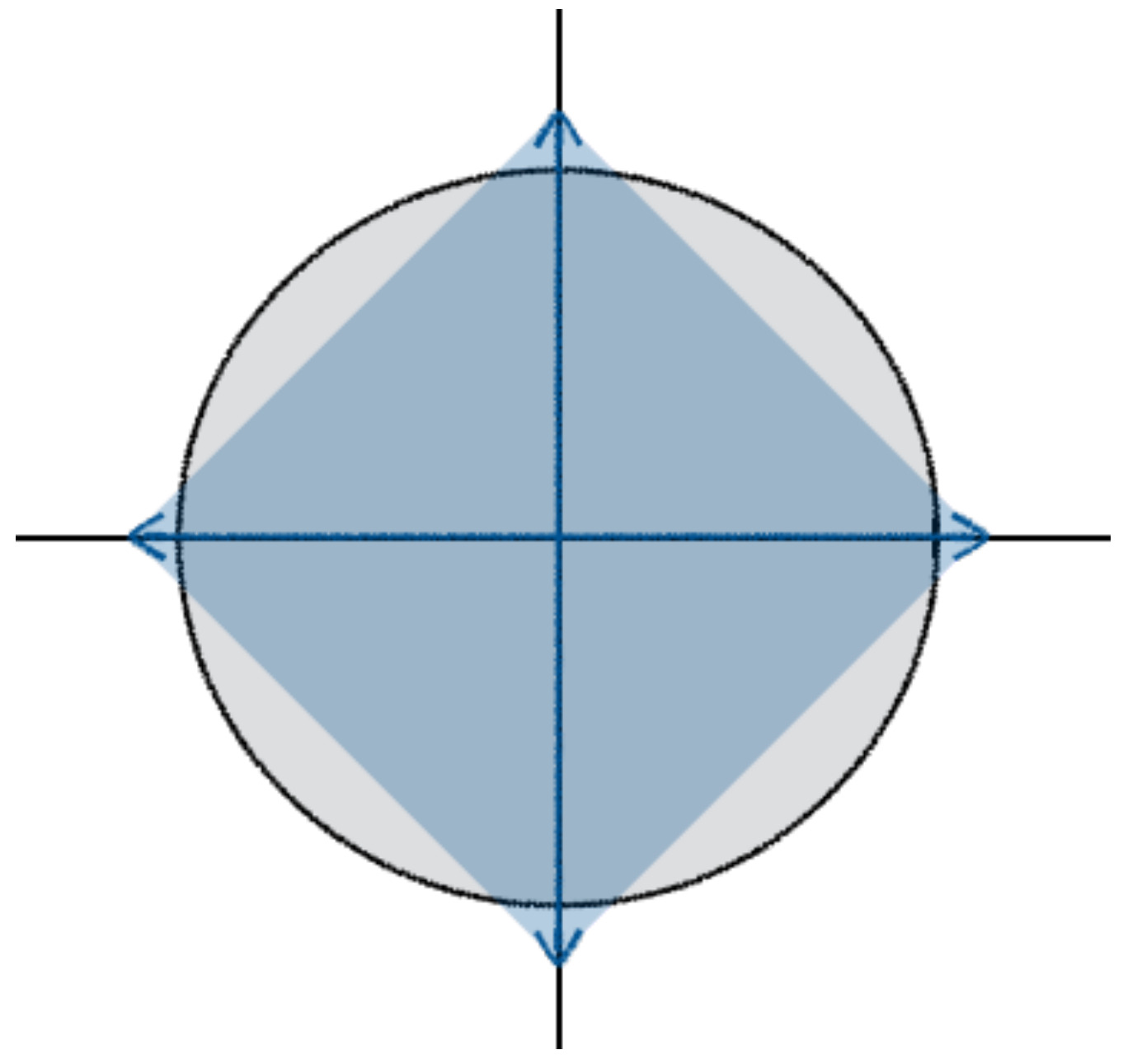}
\hspace{2cm}
\includegraphics[width=200pt]{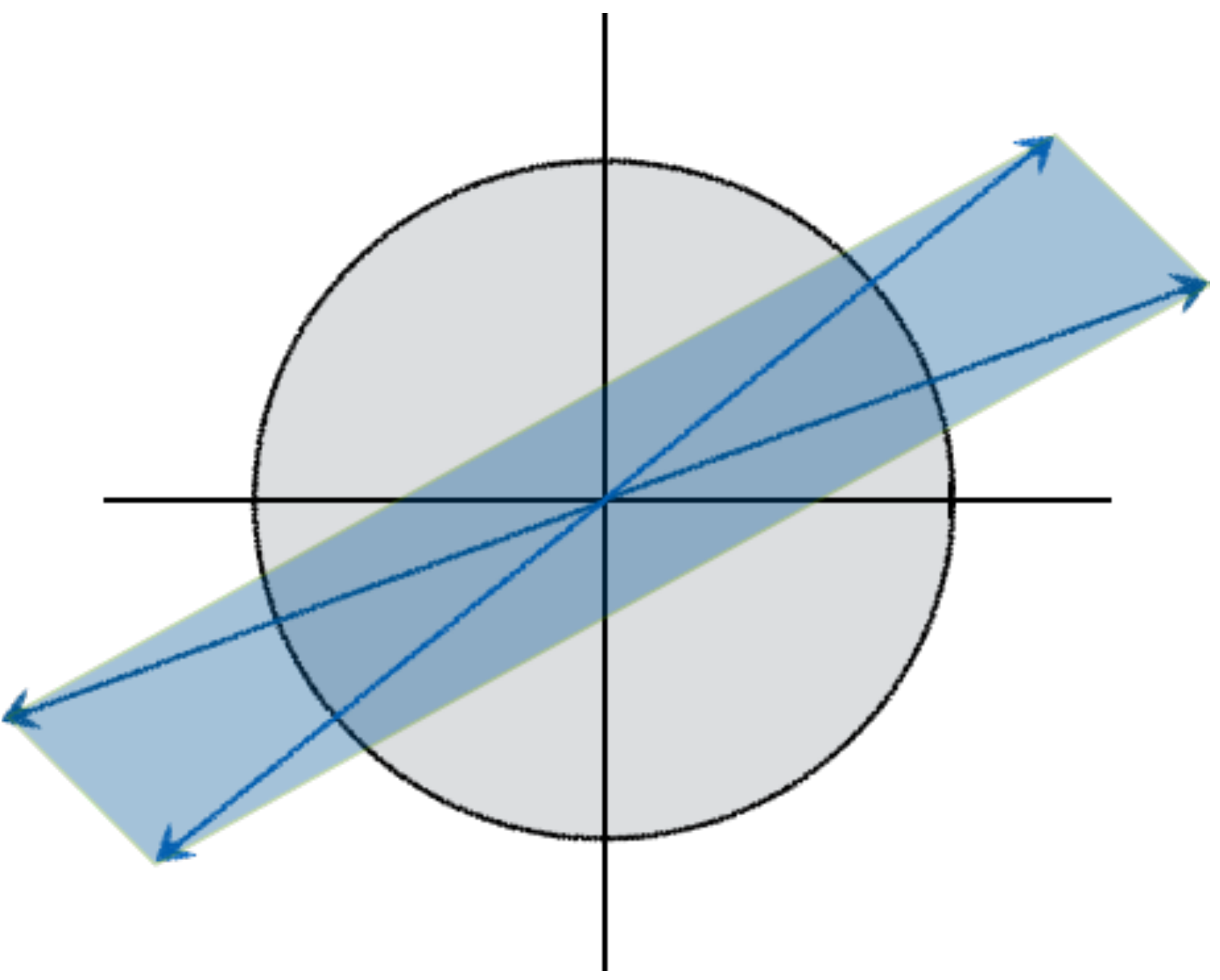}
\caption{\small Two situations in which the convex hull condition is not satisfied. Black holes within the unit disk but outside of the convex hull of the two particles (blue areas) are stable and lead to an infinite number of remnants.}\label{fig:convexhullviolation}
\end{figure}
\subsection{Generaliztion to $p$-form symmetries}

It was suggested in~\cite{ArkaniHamed:2006dz} that the WGC could be generalized to the case in which $p$-dimensional objects (e.g. $D_{p-1}$ branes in string theory) couple to p-form gauge fields. The conjecture would imply that in theories of quantum gravity, there should exist in the spectrum at least one charged object whose tension to charge ratio satisfies
\be
\frac{T}{gM_p}\leq 1
\ee
where $g$ is the charge density and $``1"$ should be defined in terms of an extremal $(p-1)$ brane.

If such a generalization holds, it would have extremely important consequences for axions and the instantons which couple to them, which corresponds to the case $p=0$. The WGC would imply that axions with large (trans-Planckian) decay constants do not arise in perturbative regimes of theories of quantum gravity. 

Although this generalization seems reasonable, and there is convincing circumstantial evidence that axions with large decay constant do not arise in string theory~\cite{Banks:2003sx,Svrcek:2006yi}, it is hard to establish a direct connection between axions and the standard form of the WGC in terms of black holes and the CEB. This connection would prove even more useful in the application of the convex hull condition to the case in which several axions mix with each other to generate a large effective decay constant. In the following sections, using T-duality, we will provide such a connection for axions that arise in the closed string sector type II theories, and show how the WGC can be used to constrain a large class of models of inflation.

\subsection{Discrete symmetries and the WGC}

Before doing that, we would like to suggest a possible further generalization of the standard WGC to the case in which the symmetries are discrete. As we will see, if such a generalization holds, it could be used to constrain models of axion monodromy inflation, which cannot be related to the standard form of the WGC.  

It is commonly assumed that discrete symmetries, just as continuous ones, should be gauged in a consistent theory of quantum gravity~\cite{Banks:2010zn}. As we mentioned above, the evidence in the continuous case is that global symmetries seem to lead unavoidably to remnant black holes with an infinite entropy, or equivalently to an infinite number of remnants only labeled by their global charge. The case against global discrete symmetries, say $\mathbb{Z}_k$, is much weaker. In particular, the entropy or equivalently the number of remnants is determined by $k$, the order of the discrete symmetry. Although for a sufficiently large $k$ one would indeed clash with the CEB, discrete symmetries of low order do not seem to be necessarily problematic. Nevertheless, it is commonly understood that all symmetries, including discrete ones of any order, are gauged. Indeed, as far as we know, all discrete symmetries found so far in string theory setups satisfy this assumption, giving circumstantial support to the conjecture.

Now, if a discrete symmetry, say $\mathbb{Z}_k$, is gauged, it must be a subgroup of a broken continuous gauge symmetry with coupling constant $g$. Just like we discussed in section~\ref{sub:conjectures}, by sending $g\to 0$ one recovers a global symmetry. If the common lore expressed in the previous paragraph is correct and discrete global symmetries are pathological, quantum gravity should prevent this limit from happening. This gives evidence that the WGC should be generalizable to the case of discrete symmetries.

In analogy with section~\ref{sub:conjectures}, imagine one has a $U(1)$ gauge symmetry with coupling constant $g$ which is broken down to a discrete $\mathbb{Z}_k$ subgroup, and a single family of particles charged under it. If the repulsive force mediated by the massive gauge bosons between two such particles is weaker than the gravitational attraction even at short distances, one could form bound states of arbitrary $U(1)$ charge. The first $k$ such states, distinguished by their discrete charge, would be exactly stable unless there is some other particle species with lower mass-to-charge ratio. For large enough $k$, one would find a conflict with the CEB. Although there is no fully convincing evidence that low values of $k$ are problematic, it is not unreasonable to conjecture that the WGC applies to discrete symmetries, even to the lowest order $k=2$. We are not aware of any counter example in string theory that contradicts this statement.

\section{Mapping instantons and axions to particles and gauge fields}
Now, we would like to verify that a restriction on large axionic field ranges does indeed follow from the standard WGC. We will do so by compactifying the theory on a circle to three dimensions, performing a T-duality and taking the decompactification limit in the dual circle. Instantons coupled  to axions in the original theory transform under this procedure into particles that couple to $U(1)$ gauge fields, to which the WGC can be applied straightforwardly.  The constraints on the spectrum of charges and masses from the WGC translate into bounds on the ratios of instanton actions and axion decay constants.

\subsection{Setup}

As an example, we consider $C_{2}$ axions in Type IIB though the procedure below works for any closed string axion.  In the conventions of \cite{polchinski1998string}, we have the following action:
\be
S_{IIB}\supset -\frac{1}{4\kappa_{0}^{2}} \int d^{10}x \sqrt{-\det g}\left(\sum_{p}|F_{p}|^{2}\right)+\mu_{p}\int_{Dp} C_{p+1}
\ee
The p-form equation of motion and charge quantization conditions are then:
\begin{eqnarray}
d*F_{p+2}&=&2\kappa_{0}^{2}\mu_{p} \delta^{9-p}(x_{\perp}) dV_{\perp} \nonumber \\
\int_{S^{8-p}} *F_{p+2} &=& 2\kappa_{0}^{2} \mu_{p}
\end{eqnarray}
We may introduce a basis of 2-forms, $w_{a}\in H^{2}(X,\mathbb{Z})$, satisfying:
\begin{eqnarray}
\int_{X} w_{a}\wedge *w_{b}= \frac{V_{X}}{\alpha'^{2}}K_{ab}
\end{eqnarray}
The matrix $K_{ab}$, the K\"ahler moduli space metric, is chosen to be dimensionless and it may encode arbitrary kinetic mixing. We now define our axion fields as:
\be
C_{2}=\sum a^{i}w_{i}
\ee
The fields $a^{i}$ will have a natural periodicity of:
\be 
\Delta a^{i}=(2\pi l_{s})^{2}
\ee
This may be understood either as a consequence of quantizing the dual 5-brane charge, or as a result of instantons.  To see how this emerges from the latter, note that the potential for $a^{i}$ arises entirely from instantons in the form of euclidean $D1$ branes wrapping a cycle $\Sigma_{k}\in H_{2}(X,\mathbb{Z})$.  An axion background will contribute a term in the action proportional to $ \mu_{1}q^{i}_{k}a^{i}$, where $q^{i}_{k}=\int_{\Sigma_{k}} w^{i}$ are integers.  This can in turn induce a non-perturbative contribution to the 4d potential of the form:
\be
V \supset \sum_{k} \Lambda^4 \, e^{-m_{k}}\left(1-\cos(\mu_{1}q^{i}_{k} a^{i})\right)
\ee
When $q^{i}_{k}=1$ for at least one 2-cycle coupling to $a^{i}$ then the periodicity is $\Delta a^{i}=2\pi/\mu_{1}=(2\pi l_{s})^{2}$ as claimed. Otherwise, the periodicity is given by:
\be
\Delta a^{i}= \frac{(2\pi l_{s})^{2}}{\gcd(q_{k}^{i})}
\ee 
The true field range, however, is only determined after we have properly normalized the kinetic terms in the action.  The relevant terms are
\beq
S_{IIB}&\supset& \frac{M_{p}^{2}}{2} \int d^{4}x \sqrt{-\det g}\left( R + \frac{g_{s}^{2}}{2(\alpha')^{2}} K_{ij} \partial a^{i} \partial a^{j} \right) \\ \nonumber
\frac{M_{p}^{2}}{2}&=& \frac{V_{X}}{(2\pi)^{7}(\alpha')^{4}g_{s}^{2}}
\eeq
Let $D^{i}_{j}$ be a matrix diagonalizing $K$, i.e, $D^{\top}KD=1$.  Then the canonically normalized fields may be written as:
\be
c^{i}=\frac{g_{s}M_{p}}{\sqrt{2}\alpha'}D^{i}_{j}a^{j}
\ee
Written in terms of these fields, the potential is now:
\be
\label{potential}
V \supset \sum_{k} \Lambda^4 \, e^{-m_{k}}\left(1-\cos\left(\mathcal{Q}^{i}_{k}\frac{c^{i}}{M_{p}}\right)\right)
\ee
where
\be
\label{Qdef}
\mathcal{Q}^{i}_{k}=\frac{1}{\sqrt{2}\pi g_{s}} (D^{-1})^{i}_{l}q^{l}_{k}
\ee
When the potential is written in the form (\ref{potential}) the field range is more easily discernible.  Unfortunately, for general $m_{k},\, \mathcal{Q}^{i}_{k}$ it is a bit difficult to define what is meant by `field range' in the first place.  If we inquire as to the range over which the potential is periodic, then it is trivial to pick a direction in field space for which the period is infinite.  A better definition would be `the range over which the slow roll conditions are satisfied'.  Sadly, this will, in general, be a quite complicated function of $m_{k}$ and we are unaware of any useful expression that encodes this definition.  One can, however, make progress by considering field range to be a concept defined purely in terms of the $\mathcal{Q}^{i}_{k}$'s, in direct analogy with the single axion case.  Following \cite{Bachlechner:2014gfa}, we thus choose a reasonable definition that is independent of $m_{k}$.  For simplicity then, set all $m_{k}=m$ and consider small perturbations around the minimum at $c^{i}=0$.  The potential is then:
\be
V\supset \frac{1}{2} \,\Lambda^4\, e^{-m} \,c \,\mathcal{Q}^{\top} \mathcal{Q} \,c
\ee
We see then that $\mathcal{Q}^{\top} \mathcal{Q}$ is proportional to the mass matrix near the origin and provides a good measure for when the potential is slowly varying.  Thus, inverse eigenvalues are a good proxy for the square of the decay constant, i.e.,
\be
\label{decayconstant}
\mathcal{Q}^{\top}\mathcal{Q} \,\tilde{c}^{(n)} = \lambda^{(n)}\, \tilde{c}^{(n)}\,\,\,\,\,  \Leftrightarrow\,\,\,\,\, f_{n}=\frac{M_{p}}{\sqrt{\lambda^{(n)}}}
\ee  
Note that this reduces to the usual definition whenever there is a single axion associated with each 2-cycle.  

\subsection{T-Duality} 
\noindent We now apply T-duality to this setup.  First, we must compactify one of the non-compact directions, say, $x^{3}$ along a circle of radius $R$.   $R$ is thus a modulus of the theory.  When $R$ takes sub-planckian values the theory is best described by $IIA$ compactified on a radius of $\tilde{R}=\alpha'/R$. In this case the $D1$ instanton maps to a massive particle represented by a $D2$ brane wrapping the same cycle.  The $C_{2}$ axions become gauge fields  with one index along $x^{3}$, from reduction of the type IIA $C_3$ form along the same cycle. 
Other polarizations of the gauge field will come from fluctuations of $C_{4}\,\,$

One might worry that Wilson lines from RR fields along the $x^{3}$ direction might lead upon T-duality to a non-isotropic background about which we know too little to reliably construct a black hole solution.  However, we may easily circumvent this problem by going to the limit where $\tilde{R}\rightarrow \infty$, in which case all anisotropies will vanish.    We may thus approach the setup with $U(1)$ gauge fields and charged scalars as studied in the original WGC.

Now our task is to relate the original parameters of type IIB to those of the final IIA theory.   Standard arguments give the IIA Planck mass ($M_{p}^{-2}=8\pi G_{4}$) and dilaton as
\beq
\label{massdilaton}
 \tilde{M}_{P}=\frac{M_{P} l_{s}}{\tilde{R}} \,\,\,\,\,\,\,\,\,\,\,\,\,\,\, \tilde{g}_{s}=\frac{g_{s} l_{s}}{R}
\eeq
For the $U(1)$'s, we work in a convention where the gauge coupling appears as part of the kinetic term and $\int A \cdot j$ appears with integral coefficient.  This convention implies
\beq
\label{guagefield}
\oint A dx^{3} \equiv \mu_{2}\int _{\Sigma_{i}\times S^{1}} C_{3}
\eeq
Equivalently,
\be
\label{guagenorm}
A^{i}_{3}=\frac{a_{i}}{(2\pi)^{2}l_{s}^{3}}
\ee
This determines the normalization of the gauge field.  Standard compactification now gives the kinetic term:
\be
S \supset \int \sqrt{-\det \tilde{g}} \frac{1}{4g^{2}} K_{ij}(F^{i})_{\mu\nu} (F^{j})^{\mu\nu}
\ee
where the parameter $g$ is:
\be
\label{charge}
g=\frac{1}{(2\pi)^{2}\tilde{M}_{P}\tilde{g}_{s}l_{s}}
\ee
Notice that the kinetic matrix $K_{ij}$ is not affected by the map we have performed, since the internal space remains untouched in the whole process.
Finally, the masses of the charged particles are simply determined by requiring the action of an instanton be equivalent to the action of a particle running in a loop.  This gives:
\be
\label{mass}
\tilde{M}_{k}=\frac{m_{k}}{2\pi \tilde{R}}
\ee
We note that $\tilde{M}_{k}$ ($m_{k}$) is by definition the exact non-perturbative mass (action) of the particle (instanton).  Thus, the arguments given here are immune to concerns regarding curvature and instanton corrections.

\subsection{Black Holes}
\label{blackhole}
Now we wish to see what constraints black holes physics in IIA places on the parameters of IIB.  Unfortunately, there is a snag.  Equation (\ref{massdilaton}) tells us that when we go to the decompactification limit in IIA we are also approaching strong coupling.  In this case, IIA gravity will no longer be a good description of the black hole solutions.  Happily, we have another useful description in this regime; M-theory.  In the M-theory description particles and gauge fields in IIA likewise lift to particles and gauge fields.  The only difference is that we are now in a 5d setting where the radius of the fifth dimension is determined by $\tilde{g}_{s}$.  In order to study possible constraints on our parameters we should therefore study 5d Reissner-Nordstrom black holes.    For the sake of generality, we will consider black holes in arbitrary space-time dimension $D=d+1$.  The lagrangian truncated to the $U(1)$ / gravity sector is:  
\begin{eqnarray}
S&=&\int d^{D}x \sqrt{-g_{D}}\left(\frac{M_{D}^{2}}{2}R_{D}-\frac{1}{4 g_{D}^{2}}K_{ij} (F^{i})^{\mu\nu}(F^{j})_{\mu\nu}\right)+\int A^{i}\wedge *j_{i}^{(5)}
\end{eqnarray}

For the case of interest, we may compactify on a circle of radius $r_{5}$ to get a relationship between the M-theory parameters and the IIA parameters:
\begin{eqnarray}
\frac{2\pi r_{5}}{g_{5}^{2}}&=&\frac{1}{g^{2}} \\ \nonumber
\frac{r_{5}}{4 G_{5}}&=&\tilde{M}_{P}^{2}=\frac{1}{8\pi \tilde{G}_{4}}
\end{eqnarray}
where we have defined $M_{D}^{-2}=8\pi G_{D}$. Now, a charged black hole has the solution:
\beq
ds^{2}&=&-h dt^{2}+h^{-1} dr^{2}+r^{2}d\Omega_{3}^{2} \\ \nonumber
h&=&1-\frac{2\tilde{M}}{(d-1) M_{D}^{2} \Omega_{d-1} r^{d-2}}+\frac{g_{D}^{2} |n|^{2}}{(d-1)(d-2)M_{D}^{2} \Omega_{d-1}^{2} r^{2d-4}} \\ \nonumber
A_{0}^{i}&=&\frac{g_{D}^{2}n^{i}}{\Omega_{d-1} r^{d-2}}
\eeq
 We have introduced the notation $|n|^{2}\equiv K_{ij}^{-1}n^{i}n^{j}$.  The general extremality condition is:
\be
\tilde{M}= g_{D}|n| \sqrt{\frac{d-1}{d-2}} M_{D}
\ee
Specializing to 5d and translating to IIA:
\be
\label{bpsbound}
\frac{\tilde{M}}{\tilde{M}_{P}} = \sqrt{\frac{3}{2}} g |n|
\ee 
We note that the extremality condition makes sense both in the M-theory language and in the IIA language.

We have managed to translate all the relevant quantities of the original type IIB theory with D1-instantons and $C_2$ axions to those of M-theory with M2-particles coupled to gauge bosons. It is worth stressing that these theories are not equivalent, they are not different descriptions of the same physical setup. Our point of view is rather that for any consistent compactification of type IIB on a given background, one should find a consistent 5d compactification of M-theory on the same internal space, with the parameters on both theories related by the dictionary just provided. If we can prove that certain parameter ranges are not allowed in any consistent compactification of M-theory, we can conclude that the translated parameters will not arise in the type IIB compactification. In this sense, we have managed to translate constraints that the standard WGC puts on 5d charged black holes to constraints on $C_2$ axions and their field ranges in 4d type IIB models. In the following, we will use this rationale to constrain models of large field inflation.

\section{WGC constraints on axions}
All of the original WGC arguments may be applied now to the 5d theory obtained by truncating M-theory to the $U(1)$ sector.  The spectrum of particles must thus obey a convex hull condition in order to prohibit the formation of an infinite number of remnants.  The size of the `unit ball' which the convex hull must contain is determined by the BPS condition obtained in (\ref{bpsbound}).  Having derived the appropriate condition in M-theory, we may now just as well use the IIA theory to discuss the resulting bound.  The IIA theory contains particles with masses given by (\ref{mass}) and charges determined as $Q^{i}_{k}=g q^{i}_{k}$ .  We may thus state the convex hull condition as:
\subsubsection*{Convex Hull Condition (IIA)}
\textit{Define the set of vectors $\vec{w}_{k}=g \vec{q}_{k}\tilde{M}_{P}/\tilde{M}_{k}$, where $(\vec{q}_{k})^{i}=q_{k}^{i}$.  Then, the convex hull generated by $\vec{w}_{k}$ must contain the ball defined by $K^{-1}_{ij}x^{i}x^{j} \le \sqrt{2/3}$.}

\noindent We now translate this to the original IIB theory using (\ref{Qdef}), (\ref{massdilaton}) and (\ref{mass}).  
\subsubsection*{Convex Hull Condition (IIB)}
\textit{Define the convex hull generated by $\vec{z}=\vec{\mathcal{Q}}_{k}/m_{k}$.  Then, the convex hull generated by $\vec{z}_{k}$ must contain the ball of radius $2/\sqrt{3}$.}

\noindent Notice that all dependence on $R$ has dropped out of the final statement.  

Although we have so far focused on axions coming from $C_2$ in type IIB theories, the results obtained above may easily be generalized to other  p-form axions.  In all cases, the strategy is to relate the axion couplings/actions to particle charges/masses in a weakly coupled theory via dualities.  In all cases we find that the moduli (e.g., the compact radius $r$) drop out of the final condition and so we arrive at a sensible condition in the original axionic variables.  In general, we may make a statement of the following form:
\subsubsection*{General Convex Hull Condition}
\label{gencond}
\textit{Let $c^{i}$ be a set of canonically normalized axions coming from a p-form compactified down to a $D=d+1$ dimensional Minkowski space.  Let the instanton couplings and actions be $\mathcal{Q}^{i}_{k}$ and $m_{k}$ respectively.  Define the convex hull, $H$, generated by $\vec{z}=\vec{\mathcal{Q}}_{k}/m_{k}$.  Then, $H$ must contain the ball of radius $r_{(p,d)}$, where $r_{(p,d)}$ is a specific $\mathcal{O}(1)$ number.}

\noindent Below, we run through some examples:
\begin{itemize}
\item The $B_{2}$ axion obeys the same constraint as the $C_{2}$ axion, being related to it via $SL(2,\mathbb{Z})$ dualities.
\item The $C_{4}$ axion in IIB dualizes to $C_{5}$ in IIA, which lifts to a 5-form in M-theory.  The instanton dualizes to a particle in IIA and to a string in the $5d$ lift.  This string behaves just like a black hole in $(3+1)d$ and so the usual BPS bound, $M \ge \sqrt{2} Q$, applies.  This leads to $r_{(4,3)}=1$.
\item Starting from a IIA axion, we may T-dualize to IIB and then apply S-duality to reach weak coupling.  One then ends up with black holes in $(3+1)d$ and so $r_{(odd,3)}=1$.  
\item More generally, an axion in IIA compactified down to $D=d+1$ dimensions is related to a IIB black hole in the same number of dimensions.  The black hole obeys a BPS bound: $M\ge \sqrt{(d-1)/(d-2)} Q$.  From this we deduce that $r_{(odd,d)}=\sqrt{2(d-2)}/\sqrt{(d-1)}$.  
\item Similarly, a $C_{2}$ axion in $d+1$ dimensions relates to a black hole in $d+2$ dimensions, so, $r_{(2,d)}=\sqrt{2(d-1)}/\sqrt{d}$.  
\end{itemize}

\subsection{General limits on axion inflation}

It is now easy to show with a few examples how naive models of effective field theories of inflation with several axions fail to satisfy the convex hull condition~\cite{Rudelius:2014wla,Rudelius:2015xta}. In fact figure~\ref{fig:convexhullviolation} illustrates two prototypical cases. The left figure corresponds to models of N-flation (for the case N=2),  while the right figure corresponds to models with aligned axions.

We now wish to give a more general proof that the the convex hull condition strictly forbids controlled axion inflation. The argument is as follows.  First, we (arbitrarily) declare our perturbative tolerance to be $e^{-\frac{1}{r_{(p,d)}}}$, or, $e^{-\sqrt{3}/2}\sim 0.421$ for the case of $C_{2}$ axions.  In other words, we don't allow any of the $m_{k}$ to sink below $1/r_{(p,d)}$.  With this constraint, we wish to show that the axion decay constant as defined by (\ref{decayconstant}) is bounded by $M_{P}$.  To simplify matters we may consider only the boundary of the region of perturbative control, i.e., $\forall k, m_{k}=1/r_{(p,d)}$.  We intend to show that \textit{even} on the boundary of the region of perturbative control one cannot obtain decay constants greater than $M_{P}$.  When the convex hull condition is satisfied, increasing $m_{k}$ only places more stringent restrictions on $\vec{\mathcal{Q}}_{k}$ and so we claim that proving the claim on the boundary is sufficient.

On the boundary of the perturbative region, the convex hull condition simplifies to the requirement that the convex hull of $\vec{\mathcal{Q}}_{k}$ contains the unit ball.  We thus need to prove that under this condition \textit{all} of the eigenvalues of $\mathcal{Q}^{\top}\mathcal{Q}$ are greater than $1$.  This will immediately imply that $f_{n}=M_{P}/\sqrt{\lambda^{(n)}}$ is sub-Planckian.  

In order to obtain the desired result, it is useful to restate the convex hull condition as follows:  \textit{The convex hull of $\vec{\mathcal{Q}}_{k}$ contains the unit ball $\iff$ for any vector $\vec{c}$ such that $|c|=1$ there is a unique set of numbers $\alpha_{k}$ and $\rho$ such that $\sum_{k}\alpha_{k}=1$, $\rho>1$, and $\sum \alpha_{k}\vec{\mathcal{Q}}_{k}=\rho\vec{c}$.}  

It is important to note that the numbers $\alpha_{k}$, $\rho$ defined above are unique.  For a square charge matrix this may be seen immediately from equation counting.  With rectangular matrices, we need a more refined argument.   Suppose, on the contrary, that the parameters $\alpha_{k}$ and $\rho$ where not unique.  We may then reach the same charge vector in two different ways, i.e.:
\be
\sum^{n} \alpha_{k}\vec{z}_{k} = \rho \vec{c},\,\,\,\,\,\,\,\,\,\, \sum^{m} \tilde{\alpha}_{j} \vec{\tilde{z}}_{j} = \tilde{\rho} \vec{c}
\ee
where $\vec{z}$ and $\vec{\tilde{z}}$ are generators of the convex hull.  Without loss of generality, we assume $\tilde{\rho} \le \rho$.  At least one of the $\tilde{\vec{z}}_{j}$ vectors must be distinct from all the $\vec{z}_{k}$'s.  Let us suppose that this is $\vec{\tilde{z}}_{m}$.  Using the equations above, we may algebraically solve for $\vec{\tilde{z}}_{m}$ by eliminating $\vec{c}$ between the two equations. Then, it is straightforward to check that $\vec{\tilde{z}}_{m}$ is actually \textit{inside} the convex hull generated by the other vectors.  (If $\rho \le \tilde{\rho}$ then we must choose $\vec{z}_{n}$ to eliminate instead.)  We can safely discard this particle from the spectrum, and the convex hull of the system will not be affected.  Proceeding in this manner, we may reach a minimal convex hull such that all vectors have a unique respresentation as stated above \footnote{The minimal convex hull need not correspond to a square matrix.}. By throwing away these extra charge vectors we are simply focusing on a certain set of instanton corrections and we will be showing that, if the convex hull contains the unit ball, large instanton corrections may be encountered already within this subset.  

Now, let us suppose that $\vec{c}$ is a normalized eigenvector of $\mathcal{Q}^{\top}\mathcal{Q}$ with eigenvalue $\lambda$.  Define $\tilde{\alpha}_{k}=\mathcal{Q}_{k}^{i}c^{i}$.  Dotting both sides of the eigenvalue equation with $\vec{c}$ one immediately finds:
\be
\label{lambdaalpha}
\lambda = \sum_{k} \tilde{\alpha}_{k}^{2}
\ee
Now, define, $\alpha_{k}$ as:
\be
\alpha_{k}=\frac{\tilde{\alpha}_{k}}{\sum_{j}\tilde{\alpha}_{j}}
\ee
These sum to one by construction.  If we now divide both sides of the eigenvalue equation by $\sum_{k}\tilde{\alpha}_{k}$ and use the equations above we get:
\be
\sum \vec{Q}_{k}\alpha_{k} = \left(\frac{\sum_{k}\tilde{\alpha}_{k}^{2}}{\sum_{k}\tilde{\alpha}_{k}}\right)\vec{c}
\ee
The uniqueness property mentioned above now implies that:
\be
\left(\frac{\sum_{k}\tilde{\alpha}_{k}^{2}}{\sum_{k}\tilde{\alpha}_{k}}\right)=\rho >1
\ee
From this it follows that at least one of the $\tilde{\alpha}_{k}$'s are greater than one, otherwise we would have a contradiction.  We thus see from (\ref{lambdaalpha}) that $\lambda$ must also be greater than one, which is what we were trying to show.  This implies that whenever we are in the regime of perturbative control, all decay constants satisfy:
\be
\label{result}
f_{n} \le M_{p}
\ee

One might ask how to interpret (\ref{result}) in light of works such as~\cite{Bachlechner:2014gfa}, which claim to give explicit realizations with $f_{n}\ge M_{p}$, in conflict with the WGC. 
If one does not give up the idea that WGC must be satisfied in string theory, we are predicting that these constructions will suffer corrections that bring them into line with the constraint~\eqref{result}. The models in fact include small (often sub-Planckian) volumes on which large corrections could arise. In particular, even if an instanton brane wrapping such a cycle does not contribute to the superpotential it can show up in the {\it scalar} potential and induce large deviations. Other sources of concern would be $KK$ and $\alpha'$ corrections, one-loop determinants and the backreaction of instanton branes, which is particularly significant in models with small cycles.

\subsection{A possible loophole}
\label{loophole}

From the above discussion it seems clear that the models with several axions so far proposed in the literature are incompatible with the WGC, both in its strong and its mild formulations. As we describe next, the mild form could still allow for a certain enhancement of the axion decay constant. It is nevertheless worth mentioning beforehand that in string theory the strong version of the WGC seems to hold, and the mechanism we discuss below has not yet found its realization in the literature.

For simplicity we will restrict our discussion to the case of a single $U(1)$ group (a single axion on the dual side), the generalization to more general cases should be simple. For the argument to hold, we need to assume that there are several families of particles (instantons) charged under the $U(1)$ group.
As we mentioned in section~\ref{sec:WGC}, the particle $(M,Q)$ that satisfies the constraint $Q \,M_p/M\geq 1$ does not need to be the lightest one in the spectrum if only the mild version of the WGC holds. Let us assume for example that there is another particle $(m,q)$, which is lighter ($M>m$), has a smaller charge, ($Q=kq$ for some $k\in\mathbb Z$), and does not satisfy the WGC bound, i.e., $q M_{p}/M < 1$. Although this situation is possible, it leads to the presence of at least $k$ stable bound states. Hence $k$ cannot be made parametrically large without running into problems with the CEB.

Translated to the T-dual language of instantons and axions, one would have two instantons coupled to a single axion. The dominant contribution comes from $(m,q)$, which is suppressed by $e^{-m}$ and has the longer periodicity $f\sim 1/q$. The other instanton has a suppression of $e^{-M} \ll e^{-m}$ and a shorter modulation $F\sim f/k$. The non-perturbative potential for the axions has the form:
\be
V= \Lambda_1^4 \, e^{-m}\left[1-\cos\left(\frac{\phi}{f}\right)\right]+ \Lambda_2^4 \, e^{-M}\left[1-\cos\left(\frac{k\phi}{f}\right)\right]\,.
\ee
The second instanton contribution is the one that ensures that the mild version of the WGC is satisfied. Nevertheless, if $m<M$, this term can be highly suppressed and negligible in comparison with the first one, which can be used to generate the inflaton potential. Notice that $m$ still needs to be bigger than $\sim 1$ to suppress higher order corrections. The requirement of trans-Planckian field range $f>M_p \Longrightarrow f\cdot m>M_p $, together with the WGC constraint satisfied by the second instanton leads to the (conservative) bound $M>m>M/k$. Since $k$ cannot be arbitrarily large, the enhancement of the decay constants cannot be made parametrically large.

This seems to be an interesting possibility by which the WGC allows for a trans-Planckian decay constant, although it only holds if the strong-WGC is violated. It is worth stressing here that this mechanism is different than any other proposed in the literature so far. The key point here is to assume that the convex hull condition is fulfilled by a set of instantons whose contributions to the potential are negligible in comparison with other instantons, which by themselves would not satisfy the WGC. Hence, for every proposal that generates a long-range inflaton potential, naively violating the WGC, one would need to ensure the existence of negligible instantons whose actions and charges satisfy the convex hull condition. This is a rather trivial step in field theory, but is highly non-trivial in explicit string embeddings. 
In fact, we are not aware of any example which violates the strong-WGC.

\section{Torsion Cycles and Generalized WGC} 

We have seen that the weak gravity conjecture prevents us from having large field ranges while keeping higher order instanton corrections under control.  These higher instantons come from multi-wrapped branes living in $H_{p}(X,\mathbb{Z})$.   One natural idea is to try and eliminate higher instantons from the spectrum by considering torsion cycles. Suppose $X$ is a 6d compact manifold with:
\begin{equation}
H_r\left(X,\mathbb{Z}\right)=\mathbb{Z}^{b_r}\oplus\mathbb{Z}_{k_1}\oplus\mathbb{Z}_{k2}\oplus...\oplus \mathbb{Z}_{k_n}
\end{equation}
Consider an instanton represened by a $D_{r-1}$ brane wrapping a torsion cycle in such a manifold.  Such an instanton cannot generate a potential for any massless closed string axion.  To see this let $\Sigma$ be a torsion cycle such that $k\Sigma=\partial\Omega$ and suppose that $w_r$ is a harmonic (hence closed) $r$-form `axion' for which we would like to generate a potential.  Then:
\begin{equation}
k\int_\Sigma w_r=\int_{k\Sigma}w_r=\int_{\partial\Omega}w_r=\int_\Omega dw_r=0
\end{equation}
Thus, $\int_\Sigma w_r=0$.  In general, the massless closed string RR field will not see the instanton at all\cite{Camara:2011jg}.  The appropriate generalization must therefore begin with massive RR fields.  This leads us naturally to the setup discussed in \cite{Marchesano:2014mla}, which provided a concrete realization of models proposed in~\cite{McAllister:2008hb,Silverstein:2008sg}.  As discussed in this reference, the general theory of torsion cycles tells us that:
\begin{equation} \label{eq:homology}
TorH_p\left(X,\mathbb{Z}\right)=TorH_{5-p}\left(X,\mathbb{Z}\right)=TorH^{p+1}\left(X,\mathbb{Z}\right)=TorH^{6-p}\left(X,\mathbb{Z}\right)
\end{equation}
Let us suppose that we have a $C_p$ RR form and let us call the fields obtained by reduction along the $C_p$ form `electric'.  The dual fields obtained by reduction of the $C_{8-p}$ form will be labelled as `magnetic'.  The cohomological relations above guarantee us a non-trivial element of $TorH^{p+1}\left(X,\mathbb{Z}\right)$.  Call this element $\omega_{p+1}$.  Then, for some $k\in\mathbb{Z}$ and some p-form $\eta_p$:
\begin{equation}
d\eta_p=k\omega_{p+1}
\end{equation}
We may now define a massive inflaton field by reducing $C_p$ along $\eta_p$:
\begin{equation}
C_p\equiv\phi\eta_p, \;\;\;\;\; F_{p+1}=d\phi\wedge\eta_p+k\phi\omega_{p+1}
\end{equation}
Dimensional reduction of the kinetic terms reproduces the monodromy model in \cite{Kaloper:2008fb,Kaloper:2011jz}.  We also have domain walls and strings in the 4d effective theory.  Again, this may be seen as a consequence of the special relations between the homology groups for torsion cycles.  A $D_{7-p}$ brane wrapped on a $TorH_{5-p}\left(X,\mathbb{Z}\right)$ cycle (which is guaranteed to exist by~\eqref{eq:homology}) will give a domain wall.  Moreover, if $\Sigma_{5-p}\in TorH_{5-p}\left(X,\mathbb{Z}\right)$, then $k\Sigma_{5-p}=\partial\Omega_{6-p}$ for some $\Omega_{6-p}$.  The $D_{7-p}$ brane wrapping this $\Omega_{6-p}$ will give a string, upon which $k$ copies of the domain wall may end.

The 4d effective potential for the axion, $\phi$, takes the form
\be
V\left(\phi\right)=\frac{1}{2}\left(q+\mu\phi\right)^2
\ee
where $\mu$ now contributes to the mass of the axion.  The parameter $q$ has dynamics sourced by the domain wall and so will jump in discrete steps across one of these domain walls.  It is also, then, quantized in terms of the brane charge $e$ such that $q=ne$ for some $n\in\mathbb{Z}$.  This gives the potential a branching structure with discrete quadratic branches labelled by $n$\cite{Kaloper:2011jz}.  

A potential problem with this scenario is that the inflaton could tunnel between different branches of the potential thus destabilizing the vacua on timescales too short to support inflation.  This tunnelling event is marked with the nucleation of a domain wall separating regions of space in different branches of the potential.  In order to maintain the monodromy, the nucleation rate of these domain walls would need to be suppressed. The tunnelling amplitude is given in terms of the tension of the domain wall, $\sigma$, and the potential difference between the two sides of the domain wall, $\Delta V$, as \cite{Coleman:1977py,Coleman:1980aw}
\begin{align}
\label{colemandeluccia}
P &\sim e^{-S_{CD}} \\ \nonumber
S_{CD} &= \frac{B_0}{\left(1+A^2\right)^2} \\ \nonumber
B_0 &= \frac{27\pi^2\sigma^4}{2\left(\Delta V\right)^3},\,\,\,\,\,\,\,\,\,\,\,\,\,\,\, A= \frac{\sqrt{3}\sigma}{\sqrt{2 \Delta V} M_{p}}
\end{align}
We parameterize the vacuum energy by $V=\beta M_{p}^{4}$ and assume that $V\sim \Delta V$.  If we impose the standard hierarchy of scales; $H < V^{1/4} \sim \Delta V^{1/4} < M_{KK} < M_{s} < M_{p}$, then one may estimate that \cite{wip}:
\be
S_{CD} \sim g_{s}^{8}\left(\frac{M_{KK}}{M_{s}}\right)^{44-4p}\left(\frac{1}{\beta}\right)^{3}
\ee
Note that $M_{KK} < M_{s}$ and so maintaining even a modest hierarchy puts a very stringent lower bound on $g_{s}$.  Thus, it seems difficult to find a parametrically large window of perturbative control where tunneling is suppressed.  To put this problem in the context of the WGC, let us suppose that WGC is strongly violated.  Notice that $\Delta V\sim q^2$, and so $A \sim \sigma/ q M_{p}$. Thus, strongly violating the WGC is equivalent to the statement that $A \gg 1$. In this regime:
\be
S_{CD} \sim \frac{1}{\beta} \gg 1
\ee
We therefore see that tunneling \textit{is} parametrically suppressed when WGC is violated and only marginally suppressed when it is satisfied.  We suspect that this statement may be made more sharp, although this would require a reassessment of the validity of the Coleman-de Luccia formula (\ref{colemandeluccia}) for non-metastable vacua  \cite{wip}.

\section{Tension with Holography}
\label{tension}
So far, we have shown that models with large axion decay constants are related through T-duality to models producing black-hole remnants.  Should there not be some sort of pathology even before performing T-duality?  In fact, such pathologies have been noted in other contexts.  In \cite{ArkaniHamed:2007js}, it was found that  constructing a certain class of Euclidean worm-holes required one to have large axionic field ranges.  Since such large field ranges are unachievable with single axions, the authors resorted to using a mechanism akin to N-flation in order to achieve the desired range.  The wormholes these authors constructed appear to be under good control within supergravity. Yet, if they really existed, these wormholes would violate the principle of locality and also cluster decomposition whenever there is a good field theory dual.  Thus, we should rigorously exclude such wormholes, even if they \textit{seem} to be under control.  We therefore see that in this context, large axion field ranges are directly tied to non-locality and may be excluded out of this
(sacred) principle.  

One might also claim that the ultimate sickness of theories with strong gravity is that they violate the covariant entropy bound (CEB) \cite{Bousso:2002ju} and that remnants are just a manifestation of this.  We can look for conflict with CEB in models with large decay constants as well.  Indeed, just such an attempt was made in \cite{Conlon:2012tz}.  In this work, the author found that in order to satisfy CEB (with no kinetic mixing) the axion decay constants must satisfy:
\be
\sum_{i}f_{i}^{2} \le \mathcal{O}(1)M_{p}^{2}
\ee
This just says that even when moving along the diagonal in axion field space, one still finds a decay constant bounded by $M_{p}$, in contrast with N-flation scenarios.  This is exactly the same result, (\ref{result}), we have found from the WGC.  It thus seems that nature's stricture against large decay constants is intimately tied to the principle of holography.   One could further argue that this provides some evidence for the strong-WGC above and beyond that which supports the mild-WGC.  If only the mild version were valid, then one could take advantage of the loophole discussed in section (\ref{loophole}) and construct inflationary theories which are in violation of the CEB.  Therefore, it seems that the CEB may actually require the strong WGC in order to be consistent.

In fact, it is not difficult to argue that not only natural inflation but generally inflation models with a large number of e-folds may be in tension with the CEB \cite{Kaloper:1999tt}.  One way to see this is as follows.  During inflation, the entropy of a region should be bounded by the corresponding `light sheet', whose area goes like $a^{2}\sim e^{2Ht}$.  However, upon reheating, we find that semiclassical physics gives an entropy proportional to the volume, or $a^{3} \sim e^{3Ht}$.  This is a parametrically large violation of CEB.  Thus, if the universe began inflating in a pure state then the observed entropy upon reheating must be due entirely to correlations entropy between local and very distant regions of the universe.  This would seem to require non-local correlations amongst particles which were never in causal contact, in direct analogy with the black hole information paradox. 

\section{Conclusions and Speculations}

In this note we have shown that many models of axion inflation are in violation of the weak gravity conjecture and are therefore unlikely to have a consistent embedding within string theory, or any other consistent theory of quantum gravity.  
The ultimate origin of the weak gravity conjecture may be found within the covariant entropy bound, which codifies the holographic nature of quantum gravity.  However, if we really take the covariant entropy bound seriously, we find tension not just with axion inflation, but with \textit{many} inflation models.

This tension would seem to indicate that there is some basic gap in our understanding of inflation, analogous to the information problem surrounding black holes. 
If we must indeed hold the covariant entropy bound sacrosanct, 
one possibility is 
to replace inflationary models with something analogous to a fuzzball structure \cite{Mathur:2005zp}, i.e, a `fuzzbang'.  In this scenario, the universe would actually satisfy the no boundary proposal \cite{PhysRevD.28.2960} and each microstate of the universe would correspond to a particular internal geometry at the time of reheating.  If this were true, then inflation would be essentially a mirage which appears to be valid in the low-energy effective field theory description.
Again, this is analogous to the fuzzball paradigm for blackholes, where the geometry may appear to be Schwarzschild outside the horizon but the interior region is simply absent.  One may also speculate along the lines of AMPS \cite{Almheiri:2012rt} that semiclassical physics breaks down upon reheating, thus thwarting one's attempt to formulate a sharp paradox with unitarity. 
  
 Thus, from the perspective of holography, the Big Bang and the Firewall appear to share some interesting similarities. 
 On the other hand, a truly explicit parametrically controlled string embedding of inflation would tell us that holography is overstepping its bounds in this setting. We have, in fact shown that there are several possible caveats which would allow one to evade some of the constraints of the WGC, e.g. if the strong or the discrete versions of the WGC are violated.  As with any ``no-go theorem'', the most interesting thing to understand is how to get around it.   Moreover, the tension between holography and large field inflation models is perhaps pointing us to a deeper understanding of quantum gravity and deserves further study.  

\subsection*{Acknowledgments}
We would like to thank Frederik Denef, Daniel Junghans, Fernando Marchesano, Thomas van Riet, Hagen Triendl, Angel Uranga and Timm Wrase for useful discussions and suggestions. We would also like thank the Michigan Center for Theoretical Physics and the organizers and participants of the workshop ``String/M-theory Compactifications and Moduli Stabilization'' where the results of this work were originally presented. This work is supported in part by the DOE grant DE-FG-02-95ER40896 and the HKRGC grants 
HUKST4/CRF/13G, 604231 and 16304414.


\bibliography{axion_v23}\bibliographystyle{utphys}

\end{document}